\documentclass[%
 reprint,
    superscriptaddress,
    amsmath,amssymb,aps,prl
]{revtex4-2}
\usepackage[normalem]{ulem}

\usepackage{graphicx}
\usepackage{dcolumn}
\usepackage{bm}
\usepackage{physics}
\usepackage{color}
\usepackage{pstricks} 
\usepackage{cancel} 
\usepackage{ulem} 
\usepackage[version=4]{mhchem}

\begin{document}

\title{Time-domain measurement of Auger electron dynamics in xenon atoms after giant resonant photoionization}

\author{Mahmudul Hasan}
\email{These authors contributed equally to this work}
\affiliation{James R. Macdonald Laboratory, Department of Physics, Kansas State University, Manhattan, Kansas, 66506, USA}

\author{Jingsong Gao}
\email{These authors contributed equally to this work}
\affiliation{James R. Macdonald Laboratory, Department of Physics, Kansas State University, Manhattan, Kansas, 66506, USA}

\author{Hao Liang}
\email{These authors contributed equally to this work}
\affiliation{Max-Planck-Institut für Kernphysik, Heidelberg, 69117, Germany}

\author{Yiming Yuan}
\affiliation{James R. Macdonald Laboratory, Department of Physics, Kansas State University, Manhattan, Kansas, 66506, USA}

\author{Zach Eisenhutt}
\affiliation{James R. Macdonald Laboratory, Department of Physics, Kansas State University, Manhattan, Kansas, 66506, USA}

\author{Ming-Shian Tsai}
\affiliation{Institute of Photonics Technologies, National Tsing Hua University, Hsinchu 300044, Taiwan}

\author{Ming-Chang Chen}
\affiliation{Institute of Photonics Technologies, National Tsing Hua University, Hsinchu 300044, Taiwan}

\author{Hans Jakob W{\"o}rner}
\affiliation{Laboratory of Physical Chemistry, ETH Z{\"u}rich, Vladimir-Prelog-Weg 2, 8093 Z{\"urich}, Switzerland}

\author{Artem Rudenko}
\affiliation{James R. Macdonald Laboratory, Department of Physics, Kansas State University, Manhattan, Kansas, 66506, USA}

\author{Meng Han} 
\email{meng9@ksu.edu}
\affiliation{James R. Macdonald Laboratory, Department of Physics, Kansas State University, Manhattan, Kansas, 66506, USA}

\date{\today}

\begin{abstract}
Time-resolved measurement of Auger-Meitner (AM) decay [Nature 419, 803 (2002)] marked a milestone in the development of attosecond science. To date, the time constants for the AM decay processes obtained from the time-domain experiments were found to be consistent with the values deduced from conventional energy-domain measurements. One of the main factors limiting the temporal resolution of these studies is the unlocked carrier-envelope-phase (CEP) of the laser pulses used to probe the electronic dynamics triggered by inner-shell photoabsorption. In this work, we report time-resolved inner-shell electron spectroscopy of xenon and krypton using attosecond soft X-ray (atto-SXR) pulses centered at 130 eV in combination with CEP-stabilized few-cycle Yb laser pulses. We observed that the N$_{4,5}$OO Auger electrons from xenon exhibit a clear streaking pattern, but with an unexpected time shift of $\sim$ 1.32 fs relative to the 4$d$ photoelectrons. Furthermore, the energy-integrated yield of streaked Auger electrons from xenon exhibits a pronounced minimum at a pump-probe time delay of 4 fs. Neither of these observations can be explained by current streaking theories and both are inconsistent with lifetimes inferred from energy-domain measurements. The M$_{4,5}$NN Auger electrons from krypton partly overlap in energy with the 3$d$ inner-shell photoelectrons and do not show these anomalous features. This study offers new insights into the inner-shell electron dynamics of heavy atoms in the giant dipole resonance region, laying the groundwork for attosecond soft X-ray spectroscopy of molecular systems containing iodine or bromine atoms.
\end{abstract}

\maketitle
The invention of attosecond light pulses and associated methodologies opened the door to time-resolving electron motion within a single atom \cite{krausz2024nobel}. An important milestone in this field was achieved in 2002 \cite{drescher2002time}, with the first time-domain study of the Auger-Meitner (AM) relaxation dynamics upon krypton M-shell ionization by attosecond pulses centered at 97 eV. In this experiment, which employed few-cycle visible pulses with random CEP to probe the timing of the Auger electron emission, the lifetime of the M-shell vacancy in krypton was found to be 7.9 fs, in good agreement with the earlier frequency-domain measurement \cite{jurvansuu2001inherent}. The development of mid-infrared driving sources for high-order harmonic generation (HHG) \cite{shan2001dramatic,popmintchev2012bright,chen2010bright,ishii2014carrier,li201753,gaumnitz2017streaking,fu2020high} and the advances in X-ray free-electron laser (XFEL) facilities \cite{duris2020tunable,guo2024experimental} has since enabled the generation of isolated attosecond pulses in the soft X-ray (SXR) regime ($>$124 eV), allowing access to faster electron dynamics in inner-shell and core levels. To date, tabletop attosecond-SXR pump-probe spectroscopy has largely relied on detecting transient absorption of optical resonances \cite{pertot2017time,attar2017femtosecond,zinchenko2021sub,yin2023femtosecond,Saito2021attosecond}, due to their higher sensitivity and energy resolution compared to charged-particle measurements. However, electron spectroscopy often provides more direct insights into the underlying ultrafast electron dynamics, particularly for inner-shell relaxation processes, as recently demonstrated using intense attosecond XFEL pulses \cite{driver2024attosecond,li2022attosecond}. Meanwhile, recent demonstrations of attosecond SXR pulses generated via HHG driven by post-compressed industrial-grade Yb lasers have shown excellent stability and high flux \cite{tsai2022nonlinear,chien2024filamentation,gao2025bright}, injecting new momentum into tabletop research on inner-shell electron dynamics.

In this Letter, we report on time-resolved inner-shell electron spectroscopy experiments of xenon and krypton atoms, with the main goal of obtaining the time-domain picture of the AM relaxation dynamics upon photoionization in the vicinity of a giant dipole resonance. AM decay is a prototypical process of inner-shell and core-level electron dynamics, where an outer-valence electron fills an inner vacancy, and the excess energy is transferred to a second valence electron, causing its ejection, as illustrated in Fig. 1(A). In the time domain, the emission of Auger electrons can be described by an exponential decay function, characterized by a constant $\tau_{\textrm{Auger}}$, known as the lifetime, which typically ranges from several to tens of femtoseconds. The lifetime of AM decay can be either deduced from the energy-domain measurements \cite{jurvansuu2001inherent} or measured using attosecond streaking techniques, employing visible \cite{drescher2002time}, mid-infrared \cite{haynes2021clocking}, or terahertz (THz) fields \cite{Schutte2012evidence} to modulate the electron momentum after the electrons transition into continuum states. However, the level of detail accessible in time-domain studies has so far been limited by the random CEP of the probe pulses used and/or by the duration of the SXR pulse. Moreover, the probe light fields not only streak the emitted electrons but also interact with the excited core-hole states in the ion \cite{Uiberacker2007attosecond,Javad2021thesis,Zhang2024time}. For the photoionization of $d$-shell electrons in xenon and krypton, the AM decay processes are particularly noteworthy due to the giant dipole resonance \cite{Ederer1964photoionization}, where strong electron-electron correlations enhance the resonant peak, extending the cross-sectional resonance over a broad energy range (80-150 eV for xenon and 100-200 eV for krypton). At present, even state-of-the-art atomic theories \cite{Pabst2013strong,Chen2015theoretical} are unable to provide \textit{ab initio} predictions for the complex, correlated electron dynamics involved.  

In this study, we used atto-SXR pulses to initiate the inner-shell dynamics, which were probed by by CEP-stabilized few-cycle Yb laser pulses (centered at 900 nm). In the time zero regime, we observed a clear energy streaking feature for the N$_{4,5}$OO Auger electrons from xenon with an unexpected time shift of $\sim$ 1.32 fs relative to the 4$d$ photoelectrons. Furthermore, long-scan time-resolved measurements reveal a yield minimum of the Auger electrons at a pump-probe delay of $\sim$ 4 fs. Both phenomena are in contradiction to the 6-8 fs Auger lifetime inferred from linewidth measurements with synchrotron radiations~\cite{carroll2002xenon,king1977investigation,kivimaki1993subnatural,masui1995new,sairanen1996high}. For krypton, the Auger electrons partially overlap spectrally with the 3$d$ photoelectrons, and thus no clear evidence of these unexpected features was observed.

Our experiments were conducted using a recently established angle-resolved streaking beamline with bright atto-SXR sources \cite{gao2025bright}. To access the giant resonance regime, we tailored an isolated atto-SXR pulse with a relatively narrow bandwidth generated via HHG from neon at a low pressure of 0.5 bar, driven by a few-cycle Yb laser and subsequently filtered through a 200-nm-thick silver filter. Figure 1(B) displays the measured HHG spectrum as a function of the CEP of the driving laser pulse, alongside the filter transmission curve. A Gaussian-shaped SXR supercontinuum spectrum, as illustrated in Fig. 1(C), spanning from 110 eV to 150 eV, appears at a relative CEP of 3.2 $\pi$ and is used to ionize xenon and krypton atoms. The duration of the attosecond pulse was characterized to be approximately 100 as based on the photoelectron streaking trace \cite{gao2025bright}. A high-energy velocity map imaging (VMI) spectrometer \cite{kling2014thick,hasan2025strong} was used to measure the photoelectron momentum distributions as a function of the time delay between the atto-SXR pulse and the short IR pulse. Further experimental details are provided in the Supplementary Materials (SM).

\begin{figure}[htbp!]
\centering
\includegraphics[width=\linewidth]{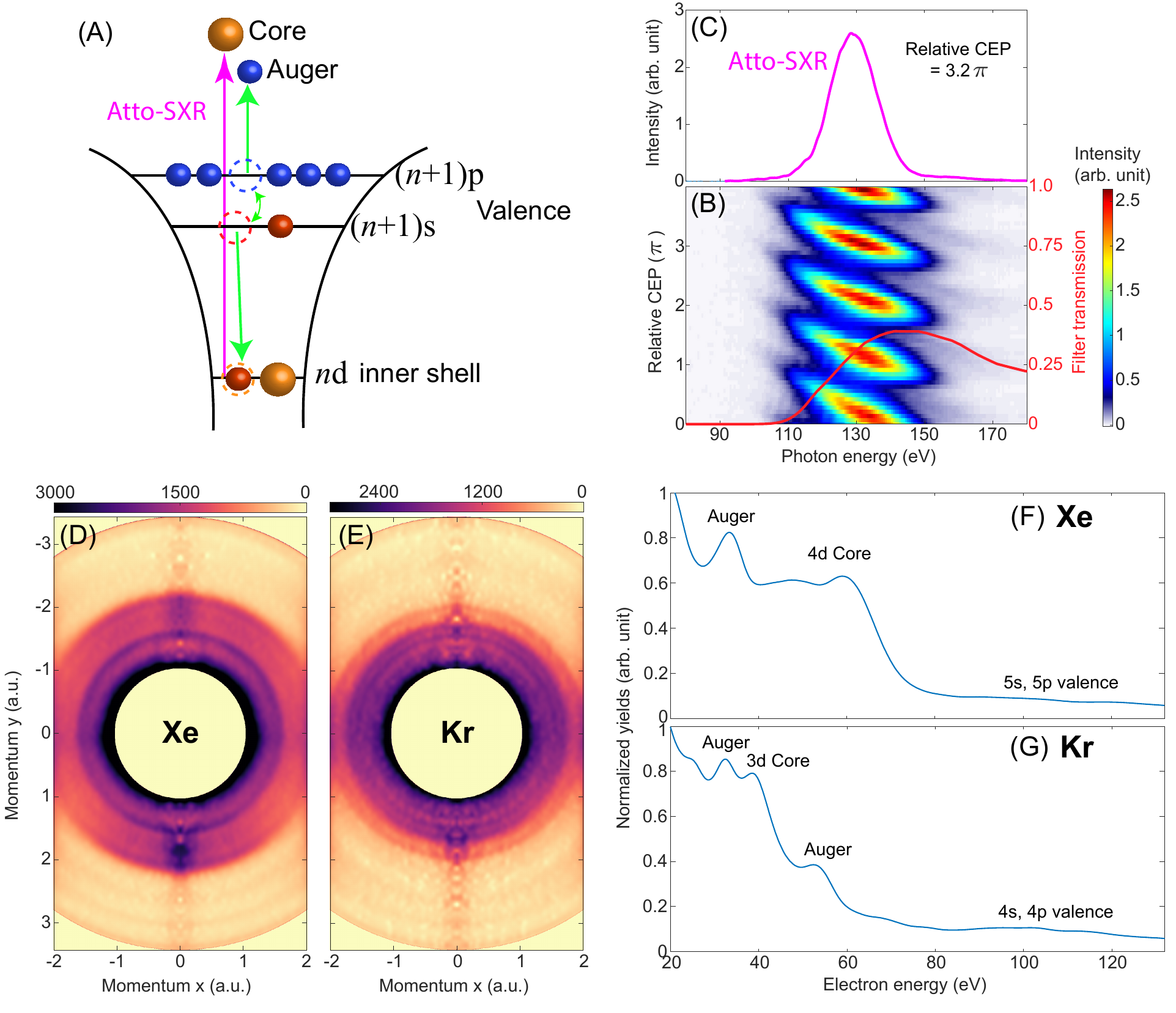}
\caption{\textbf{Inner-shell electron spectroscopy of xenon and krypton atoms ionized by tabletop attosecond soft X-ray pulses.} (\textbf{A}) Schematic diagram of photoelectron (magenta arrow) and Auger electron (green arrow) emissions. (\textbf{B}) CEP-resolved  HHG supercontinuum spectra measured after a 200-nm-thick Ag filter, where the filter transmission (red curve) is overlaid. (\textbf{C}) Very smooth Gaussian-shaped HHG spectrum at the relative CEP of 3.2 $\pi$, which is used in the ionization experiments. (\textbf{D, E}) Measured electron momentum distributions from xenon and krypton atoms, respectively. The distributions are Abel-inverted central-momentum slices and thus there are some artifacts along the vertical axis. (\textbf{F, G}) Corresponding electron energy distributions after an angular integration from 5$^\circ$ to 30$^\circ$, where 0$^\circ$ is defined as the vertical axis (i.e. polarization direction).}
\label{fig:figure1}
\end{figure}

Figures 1(D) and 1(E) show the measured electron momentum distributions after Abel inversion for the atto-SXR pulse only, obtained from xenon and krypton atoms, respectively. A series of concentric rings are observed. Figures 1(F) and 1(G) present the corresponding energy spectra, providing a clearer assignment of the channel contributions. The outermost rings at around 100 eV correspond to photoelectrons originating from the valence orbitals with small cross sections in the soft X-ray regime, whereas the stronger inner rings are corresponding to the inner-shell photoelectrons and Auger electrons. In xenon, the 4$d$ photoelectrons are centered at 58 eV, followed by an Auger electron band at 33 eV. The 33-eV Auger band is corresponding to the channel decayed to the ground state ($5p^{-2}$) of Xe$^{2+}$ (see more discussion on Figure 4). In krypton, the 3$d$ photoelectrons are centered at 40 eV, situated between two Auger electron bands at 54 eV and 32 eV, corresponding to the $4p^{-2}$ and satellite $4p^{-3}nl$ states of Kr$^{2+}$, respectively. These assignments are both in good agreement with previous synchrotron measurements of xenon \cite{carroll2002xenon} and krypton \cite{schmidtke2001kr}.

\begin{figure}[htbp]
\centering
\includegraphics[width=\linewidth]{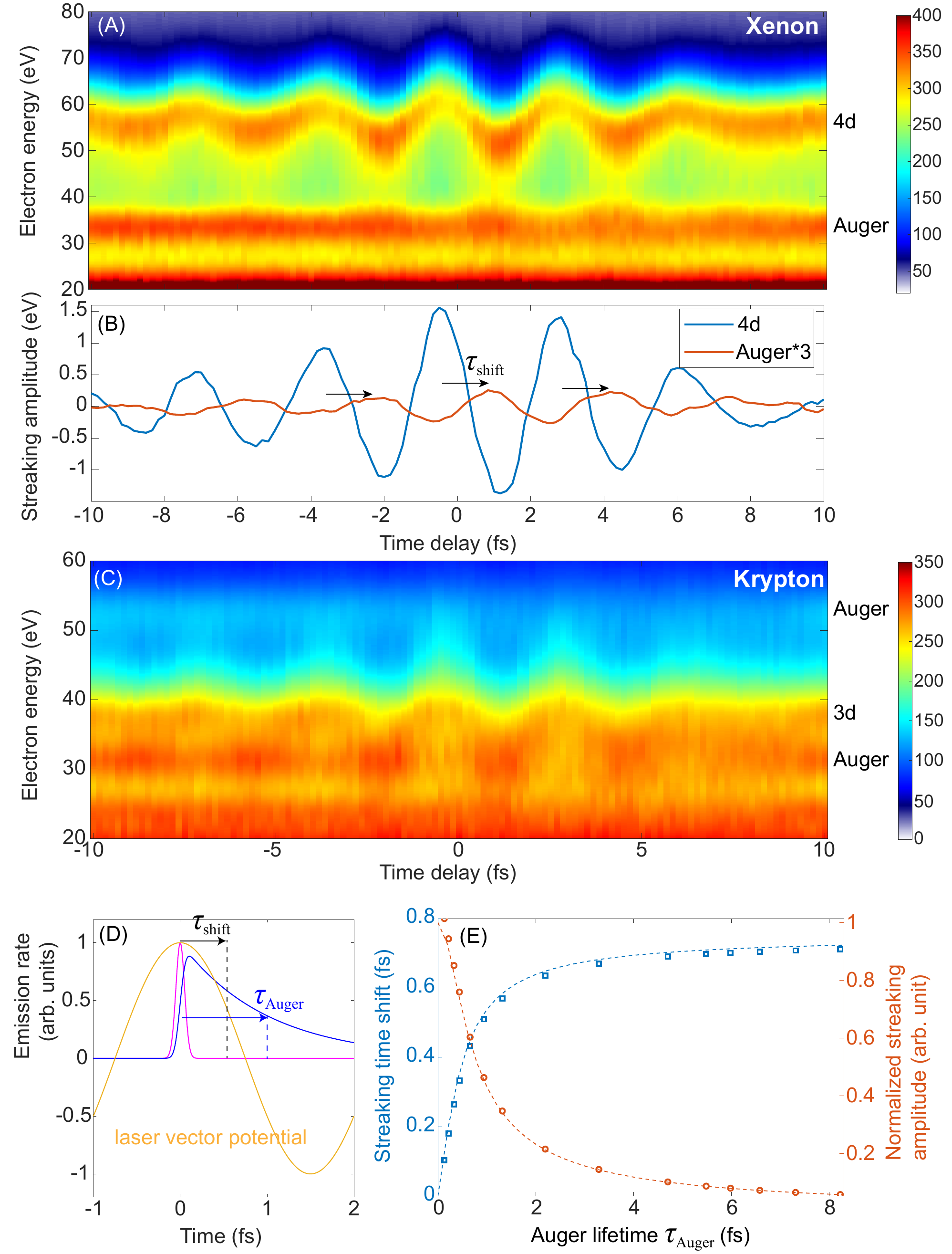}
\caption{\textbf{Attosecond streaking measurements on xenon and krypton atoms.} (\textbf{A, C}) Measured electron streaking traces with an angular integration from 5$^\circ$ to 30$^\circ$ from xenon and krypton, respectively. In (A) we observe the momentum streaking effect on both 4$d$ photoelectrons and Auger electrons with a time shift $\tau_{\rm{shift}}$. To quantify this time shift, (\textbf{B}) shows the extracted streaking amplitudes with a magnification for Auger electrons. (\textbf{D}) Schematic illustration of the streaking effect on photoelectrons (magenta curve) and Auger electrons (blue curve). The generation rate of photoelectrons is proportional to the intensity envelope of the attosecond SXR pulse, while the generation of Auger electrons is modeled as a convolution with an exponential decay function, reflecting the finite lifetime of the Auger-Meitner process. (\textbf{E}) Calculated streaking time shifts and streaking amplitudes of the Auger electrons as a function of the their lifetime. The dashed lines represent the classical analytical results [Eq. (1)] and the data points are the results from quantum simulations. The predicted largest time shift is approaching to 0.75 fs, i.e., a quarter of an optical cycle. However, our experimental data reveal a time shift of approximately 1.32 fs, posing a challenge to current theories.}
\label{fig:figure2}
\end{figure}

Our first time-resolved experiment consists of attosecond streaking measurements in which a CEP-stabilized IR pulse is temporally overlapped with the atto-SXR pulse. Electrons emitted at different instants can be distinguished by the distinct streaked momenta they acquire from the IR field. Figure 2(A) presents the photoelectron energy spectrogram of xenon as a function of the time delay between the two pulses, which are emitted along one side of the polarization direction. A pronounced energy oscillation (i.e., streaking phenomenon) is observed for the 4$d$ photoelectron band, while a weaker yet discernible streaking phenomenon appears for the Auger band, exhibiting a substantial time shift compared to the 4$d$ band. To quantify this shift, Figure 2(B) displays the streaking amplitudes for the two bands, with a subtraction to their delay averaged values. The streaking amplitude is defined as the weighted energy, i.e. $\int W(\tau,E_k)E_kdE_k/\int W(\tau,E_k)dE_k$ with the photoelectron yield $W(\tau,E_k)$ at the time delay of $\tau$ and the energy of $E_k$. A magnification factor of $3$ is applied to that of Auger band. The streaking time shift $\tau_{\rm{shift}}$ between the photoelectron and Auger electron emissions, determined from the streaking amplitudes, is approximately 1.32 fs, close to the half optical cycle of the IR pulse, as indicated by the black arrows in Fig. 2(B). In the SM, we verify that the 4$d$ and valence photoelectron streaking traces are almost synchronized. The 4$d$ electrons therefore serve as the timing reference (the IR vector potential), as the residual photoionization delay is negligible. We additionally provide two videos of the angle-resolved streaking for both raw VMI images and Abel-inverted ones.

The pronounced 1.32-fs time shift observed for Auger electrons cannot be explained by the streaking picture based on the lifetime of approximately 6-8 fs inferred from synchrotron measurement~\cite{carroll2002xenon,king1977investigation,kivimaki1993subnatural,masui1995new,sairanen1996high} and 6 fs from a recent extreme ultraviolet four-wave mixing measurement~\cite{puskar_probing_2025}. Figure 2(D) displays a schematic illustration of the streaking principle on photoelectrons (magenta curve) and Auger electrons (blue curve). The emission rate of photoelectrons follows the intensity envelope of the atto-SXR pulse, whereas the emission of Auger electrons is modeled as a convolution with an exponential decay function, accounting for the finite lifetime $\tau_{\textrm{Auger}}$ of the AM process. From a classical perspective, the Auger electron has an emission probability of $\exp(-t/\tau_{\textrm{Auger}})\,\Theta(t)/\tau_{\textrm{Auger}}$ at time $t$, after which it is streaked by the IR-field vector potential $A(t)=A_{0}\cos[\omega(t-\tau)]$, where $\Theta(t)$ denotes the Heaviside step function. Thus the averaged momentum shift of the Auger electron is
\begin{equation}
    \begin{aligned}
       \langle\Delta p(\tau)\rangle=\int_0^\infty A_0\cos[\omega(t-\tau)]\exp(-t/\tau_{\textrm{Auger}})/\tau_{\textrm{Auger}}\,\mathrm dt \\
       =\frac{A_0}{\sqrt{1+(\omega \tau_{\textrm{Auger}})^2}}\cos[\omega\tau-\arctan(\omega \tau_{\textrm{Auger}})].
    \end{aligned}
\end{equation}
For a short Auger lifetime, the streaking amplitude of Auger electron is roughly the same as that of photoelectron, and the streaking time delay is approximately equal to Auger lifetime.
A longer Auger lifetime leads to a reduced streaking amplitude with the ratio of $1/\sqrt{1+(\omega \tau_{\textrm{Auger}})^2}$ and a larger time shift of $\arctan(\omega \tau_{\textrm{Auger}})/\omega$ with respect to the vector potential. When the Auger lifetime exceeds one optical cycle, the streaking effect becomes much weaker, and multi-cycle interference–induced sidebands emerge \cite{drescher2002time}. Notably, for a very long AM decay, the maximum time shift of the Auger electrons approaches a quarter of the optical cycle ($\arctan(\infty) = \pi/2 \approx 0.75\,\rm{fs}$), as the electrons effectively acquire little to no net energy from the laser field.

To verify the above analysis, we followed the pioneer work done by Smirnova {\it et al.} \cite{smirnova_quantum_2003}, simulated the doubly ionization AM process. We have adopted the strong field approximation, ignoring electron-ion and electron-electron interaction after ionization, and the central momentum approximation, regarding the ionization dipole and the configuration interaction as energy independent constants (see SM for further theoretical details). After calculating the double-ionization spectra for different lifetimes (see SM for several representative streaking traces), we extracted the corresponding streaking time shifts and amplitudes using the same analysis procedure as for the experimental data. The numerical results are shown in Fig.~2(E), together with the prediction of the classical model. Although this classical model does not include sidebands or intercycle interference, it accurately reproduces the delay dependence of the weighted central momentum, in good agreement with the quantum simulations. As the Auger lifetime increases, the streaking amplitude decreases, while the time shift increases and eventually saturates at approximately $0.75~\mathrm{fs}$. 

The Auger lifetime can, in principle, be inferred from either the streaking amplitude or the time shift. However, these two approaches yield mutually inconsistent results. The measured ratio between the streaking amplitudes of the photoelectrons and the Auger electrons is $14.3 \pm 0.8$, corresponding to an estimated lifetime of $6.8 \pm 0.4 \ \mathrm{fs}$, which is in reasonable agreement with the literature \cite{carroll2002xenon,puskar_probing_2025}. In contrast, the experimentally observed time shift of roughly 1.32 fs exceeds the largest value predicted by theory, posing a challenge to current theoretical understanding.

\begin{figure}[htbp]
\centering
\includegraphics[width=\columnwidth]{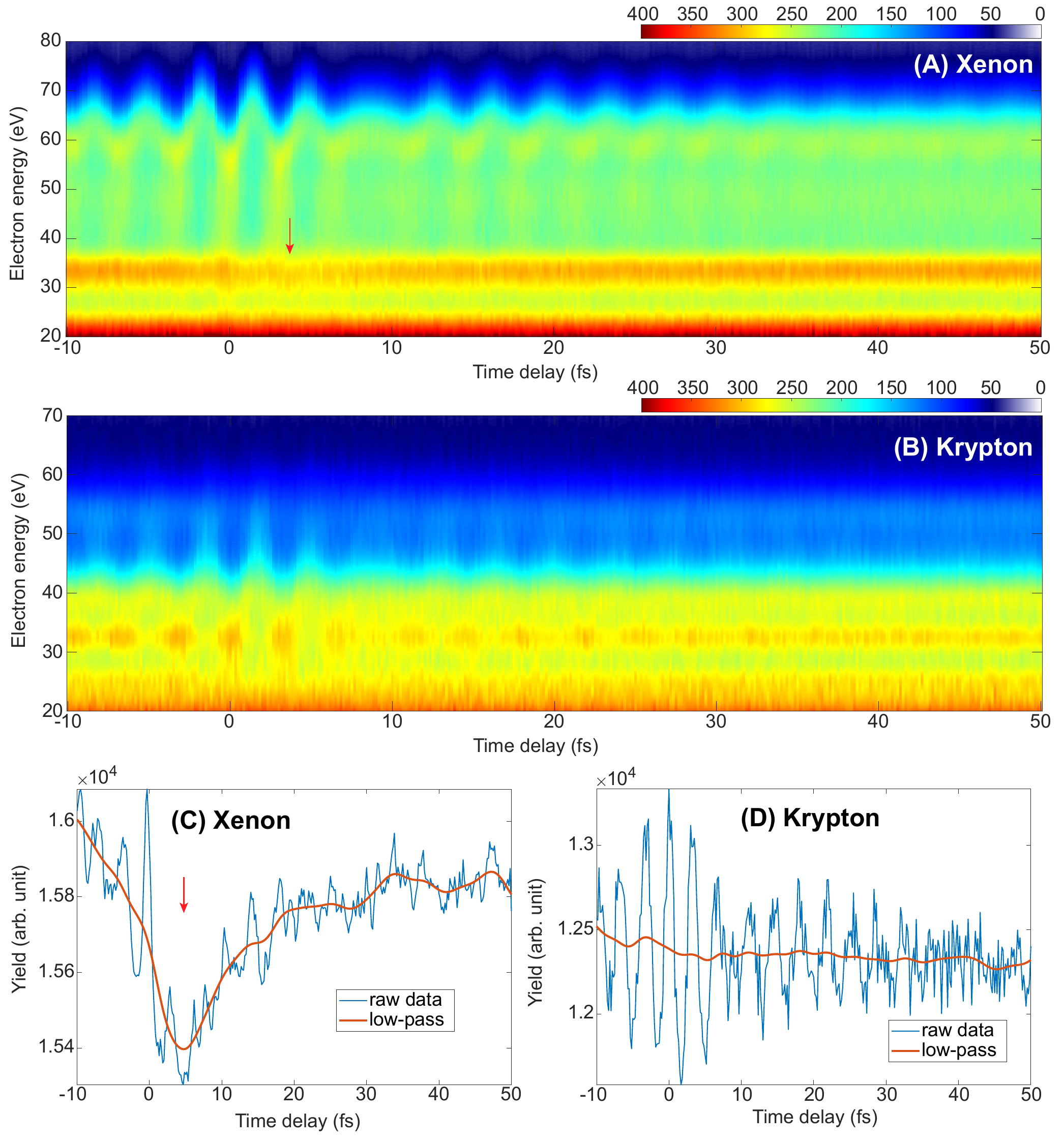}
\caption{\textbf{Long-range pump-probe experimental data}. (\textbf{A-B}) Measured electron energy spectra as a function of pump-probe time delay for Xe and Kr atoms, respectively. (\textbf{C-D}) Energy-integrated Auger electron yields in the ranges of 29–35 eV for krypton and 29–37 eV for xenon. The orange curves represent the results after applying a low-pass filter ($<$0.5 eV). In (A) and (C) the red arrow marks the delay position of the yield minimum of Auger electrons.}
\label{fig:figure3}
\end{figure}

Our second measurement is a pump–probe experiment in which the time delay between the two pulses is scanned from the overlapping to the well-separated regime. Figures 3(A,B) show the time-resolved electron energy spectra for xenon and krypton, respectively. For xenon, we observe that the energy-integrated electron yield of the whole Auger band centered at 33 eV reaches a minimum at a delay of 4 fs, as indicated by the red arrow in Fig. 3(A). The corresponding raw electron yield and the result after applying a low-pass frequency filter are shown in Fig. 3(C). In Fig. 3(D), we present the Auger-band result for krypton over the similar energy range. There, a rapid yield oscillation arises from leakage from the $3d$ photoelectron band. After applying the low-pass filter, the Auger yield becomes nearly flat, consistent with the general streaking picture in which the IR field modulates the energy distribution without altering the total yield. The energy-integrated photoelectron yield suggests this general picture. The minimum in the xenon Auger yield at 4 fs therefore suggests that additional, faster decay channels may be enabled by the IR pulse and temporarily dominate over the normal AM process in the delay window from 0 to 4 fs. 

\begin{figure}
    \centering
    \includegraphics[width=\columnwidth]{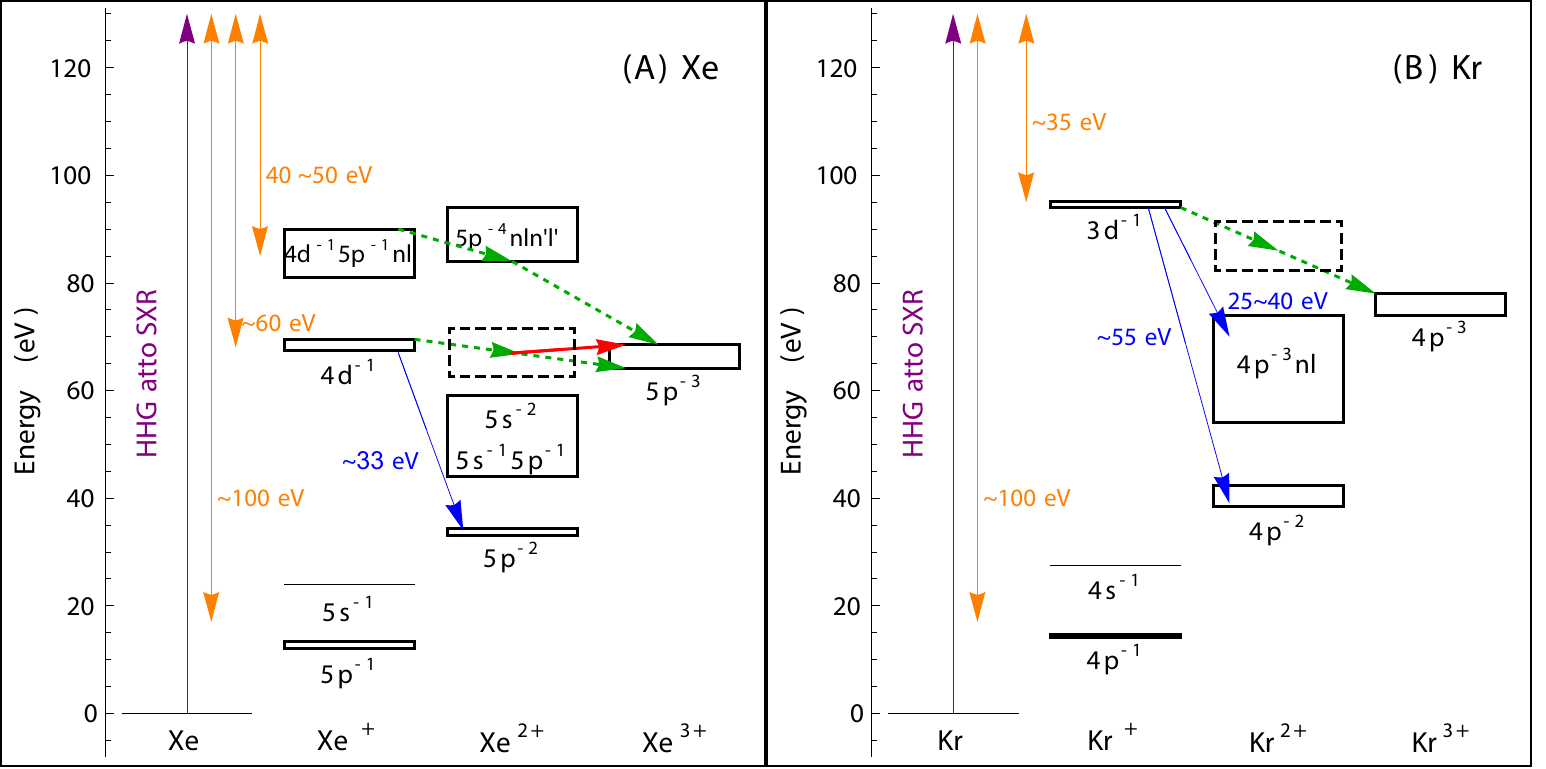}
    \caption{Related energy levels and transitions for (A) Xe~\cite{penent_multielectron_2005} and (B) Kr~\cite{palaudoux_multielectron_2010}. The dashed boxes indicate a manifold of Rydberg satellite states.}
    \label{fig:figure4}
\end{figure}

In Figure 4(A,B) we show the binding energies of related levels and transitions among singly, doubly and triple charged ions, and the energy gap between two states is corresponding to the kinetic energy of an emitted electron. For Xe, the 33-eV Auger band is corresponding to the transition from the $4d^{-1}$ hole state to the ground state of Xe$^{2+}$, i.e., $5p^{-2}$. The energy gap between $5p^{-2}$ and $5s^{-1}5p^{-1}$ or $5s^{-2}$ is larger than 10 eV. In our experiments, the IR intensity is accurately calibrated at $4\times10^{11}\ \mathrm{W/cm^{2}}$ according to the streaking amplitude of photoelectrons, which is too low to significantly couple the $5p^{-2}$ state to the higher-lying $5s^{-1}5p^{-1}$ or $5s^{-2}$ states. Consequently, a reduction of the Auger yield at 33~eV is unlikely. Another scenario is that the IR field is able to tunnel ionize high Rydberg electrons in the $5p^{-3}nl$ states, which was found in previous studies \cite{Uiberacker2007attosecond,Javad2021thesis,Zhang2024time}, and therefore this pathway ($4d^{-1} \rightarrow5p^{-3}nl\xrightarrow{\rm{IR}}5p^{-3}$) is expected to dominate over than the channel of our interest ($4d^{-1} \xrightarrow{33~\rm{eV}}5p^{-2}$). However, this scenario still cannot explain the yield minimum at 4 fs, since the lifetimes of the $5p^{-3}nl$ states is typically longer than 6 fs \cite{puskar_probing_2025}. Such long lifetimes cannot produce a yield minimum at $4~\mathrm{fs}$, even without accounting for the convolution effect by the IR pulse duration. Our intriguing and unexpected observations therefore call for more in-depth future investigations.

In conclusion, we have performed attosecond soft–X-ray pump–probe spectroscopy of xenon and krypton atoms and uncovered two unexpected features in the N$_{4,5}$OO Auger-Meitner decay of xenon: a large streaking time shift of $\sim$1.32~fs relative to the 4$d$ photoelectrons and a pronounced minimum in the integrated Auger yield at a delay of $4$~fs. Both observations lie far outside the predictions of established streaking models and are inconsistent with lifetimes obtained from energy-domain measurements, indicating that additional, currently unresolved many-electron or strong-field interactions influence the decay dynamics in the giant dipole resonance region. These results demonstrate previously unrecognized complexity in inner-shell electron dynamics of heavy atoms and pave the way for attosecond SXR studies of molecular systems containing heavy elements.

We thank C. Aikens, J. Millette and S. Chainey for their technical support. M. Han thanks Prof. Sheng Meng for discussions. The Kansas group was supported by the Chemical Sciences, Geosciences and Biosciences Division, Office of Basic Energy Sciences, Office of Science, US Department of Energy, Grant No. DE-FG02-86ER13491. M.-C.C. thanks the National Science and Technology Council, Taiwan, for funding grant no. 113-2112-M-007-042-MY3.

\bibliography{pop_references}

\end{document}